\documentclass[twocolumn,showpacs,preprintnumbers,amsmath,amssymb,APSl,prd,nofootinbib,superscriptaddress]{revtex4-1}

\usepackage{dcolumn}
\usepackage{bm}
\usepackage{ifpdf}
\usepackage{hyperref}
\usepackage{dcolumn}
\usepackage{bm}
\usepackage[spanish,english]{babel}
\usepackage{amsfonts}
\usepackage{amssymb}
\usepackage{graphicx}
\usepackage[latin1]{inputenc}

\usepackage{amsmath}
\usepackage{tikz}
\usetikzlibrary{decorations.pathmorphing}

\newcommand{\be}{\begin{equation}}
\newcommand{\en}{\end{equation}}
\newcommand{\bea}{\begin{eqnarray}}
\newcommand{\ena}{\end{eqnarray}}

\begin{document}
\title{Classical resolution of black hole singularities via wormholes}

\author{Gonzalo J. Olmo} \email{gonzalo.olmo@csic.es}
\affiliation{Departamento de F\'{i}sica Te\'{o}rica and IFIC, Centro Mixto Universidad de
Valencia - CSIC. Universidad de Valencia, Burjassot-46100, Valencia, Spain}
\affiliation{Departamento de F\'isica, Universidade Federal da
Para\'\i ba, 58051-900 Jo\~ao Pessoa, Para\'\i ba, Brazil}
\author{D. Rubiera-Garcia} \email{drgarcia@fc.ul.pt}
\affiliation{Instituto de Astrof\'isica e Ci\^encias do Espa\c{c}o, Universidade de Lisboa, Faculdade de Ci\^encias, Campo Grande, PT1749-016 Lisboa, Portugal}
\affiliation{Center for Field Theory and Particle Physics and Department of Physics, Fudan University, 220 Handan Road, 200433 Shanghai, China}
\author{A. Sanchez-Puente} \email{asanchez@ific.uv.es}
\affiliation{Departamento de F\'{i}sica Te\'{o}rica and IFIC, Centro Mixto Universidad de
Valencia - CSIC. Universidad de Valencia, Burjassot-46100, Valencia, Spain}

\date{\today}

\begin{abstract}
In certain extensions of General Relativity, wormholes generated by spherically symmetric electric fields can resolve black hole singularities without necessarily removing curvature divergences. This is shown by studying geodesic completeness, the behavior of time-like congruences going through the divergent region, and by means of scattering of waves off the wormhole. This provides an example of the logical independence between curvature divergences and space-time singularities, concepts very often identified with each other in the literature.
\end{abstract}


\maketitle

\section{Introduction}

The blow up of curvature scalars or of components of the Riemann tensor is typically used in the literature to put forward the existence of space-time singularities. Indeed, it is common to use those divergences to argue that when curvature scalars get close or surpass Planckian scales it is time to replace Einstein's theory by an improved description, opening a door to quantum theories of gravity. This intuitive view has even shaped numerous approaches in the search of nonsingular space-times by trying to build theories with bounded curvature scalars \cite{Mukhanov:1991zn, Ansoldi:2008jw,Lemos, Spallucci, Bronnikov,Hayward}. However, from a formal perspective, the most widely accepted criterion to characterize a singular space-time is through the existence of incomplete paths (inextendible time-like or null geodesics) \cite{Geroch:1968ut,Hawking:1973uf,Wald:1984rg} (see also \cite{Curiel2009} and references therein).

Given that geodesics are classical geometrical entities associated with idealized (structureless) physical observers, it is unclear what a quantum theory of gravity should say about them. The focus is thus typically turned back to curvature divergences as one of the disturbing elements an improved theory of gravity should get rid of. A combined approach to better understand the correlation observed between the incompleteness of geodesics and the appearance of curvature divergences consists on exploring the impact of curvature divergences on physical observers represented by geodesic congruences. This has led to a classification of the strength of curvature divergences \cite{Ellis, Tipler, CK, Nolan, Ori} according to whether an object going through them is crushed to zero volume or ripped apart by infinite tidal forces ({\it strong} case), or if  it is not severely affected ({\it weak} case).  In the end, however, one must accept that, to the best of our knowledge, matter and energy have quantum properties and a fundamental wave-like behavior, which demands for a characterization of curvature divergences and space-time singularities by means of quantum scattering experiments \cite{Giveon:2004yg}.

In this work we provide an example of black hole space-time in which a wormhole gets rid of the central singularity without necessarily avoiding curvature divergences. This geometry is an exact electrovacuum solution of certain high-energy extensions of general relativity (GR) formulated in metric-affine spaces (see \cite{Olmo:2012nx,Lobo:2013adx,ORGS14}, where this geometry is derived in detail). The requirement of a metric-affine geometry is essential to avoid ghosts and higher-order derivative equations. The resulting space-time roughly consists on two copies of the Reissner-Nordstr\"{o}m or Schwarzschild black hole solutions\footnote{Depending on particular parameters, two copies of Minkowski are also possible.} connected by a wormhole in a small region near the center. This wormhole is supported by an electric field and, as such, satisfies all the classical energy conditions.

Incomplete geodesics that in GR end (or start) at the central singularity can now go through the wormhole, thus avoiding incompleteness \cite{ORGSPa}. The wormhole throat, however, generically exhibits divergent curvature scalars. Nonetheless, we find that physical observers (described as time-like congruences) do not experience any pathological behavior upon crossing the wormhole \cite{Olmo:2016fuc}. We also find that a scalar field propagating in this background is everywhere well-behaved, and compute the transmission coefficients and cross section of scalar waves as they interact with the wormhole \cite{ORGSPc}. The resolution of black hole singularities in this model provides new insights on how this disturbing aspect of classical gravitation may be cured at high energies.

\section{Background geometry and geodesics}

The geometry we are going to study arises when a static, spherically symmetric electric field is coupled to a certain extension of GR (see \cite{Olmo:2012nx,ORGS14} for details). In fact, this solution appears in two different gravity theories, namely,  the quadratic gravity model with Lagrangian $R-\frac{\epsilon}{2}(R_{\mu\nu}R^{\mu\nu}+aR^2)$ and the Born-Infeld gravity model $\frac{1}{\kappa^2\epsilon}(\sqrt{-|g_{\mu\nu}+\epsilon R_{\mu\nu}|}- \sqrt{-g})$. The line element, whose derivation was worked out in \cite{Olmo:2012nx,ORGS14},  takes the form
\begin{equation}\label{eq:ds2}
ds^2=-A(r)dt^2+\frac{1}{B(r)}dx^2+r^2(x)d\Omega^2 \ ,
\end{equation}
with $d\Omega^2=d\theta^2+\sin\theta^2 d\varphi^2$, $B(r)=A(r) \sigma_{+}^2$, and
\begin{eqnarray}
r^2(x)&=&\frac{x^2+\sqrt{x^4+4r_c^4}}{2} \label{eq:WH} \\
A(x)&=& \frac{1}{\sigma_+}\left[1-\frac{r_S}{ r  }\frac{(1+\delta_1 G(r))}{\sigma_-^{1/2}}\right] \\
\sigma_\pm&=&1\pm \frac{r_c^4}{r^4(x)}
\end{eqnarray}
where $x\in ]-\infty,+\infty[$, but $r(x)\ge r_c>0$. The function $G(z)$, with $z=r/r_c$, can be written as
\begin{equation}
G(z)=-\frac{1}{\delta_c}+\frac{1}{2}\sqrt{z^4-1}\left[f_{3/4}(z)+f_{7/4}(z)\right] \ ,
\end{equation}
where $f_\lambda(z)={_2}F_1 [\frac{1}{2},\lambda,\frac{3}{2},1-z^4]$ is a hypergeometric function, and $\delta_c\approx 0.572069$ is a constant. By careful inspection, one can verify that this line element describes a wormhole \cite{Visser}. To see this, Eq.(\ref{eq:WH}) is particularly useful, as it shows that the area function $A=4\pi r^2(x)$ has a minimum $A_{min}=4\pi r_c^2$ at $x=0$, where the wormhole throat is located. Here $r_c\equiv \sqrt{r_q l_\epsilon}$,  $l_\epsilon$ is a length scale characterizing the departures from GR \cite{Olmo:2012nx,ORGS14} (in fact $\epsilon=-2l_\epsilon^2$),  $r_q^2\equiv 2G q^2$ is the charge radius, $r_S$  the Schwarzschild radius, and $\delta_1\equiv (2r_S)^{-1}\sqrt{r_c^3/l_\epsilon}$ is the charge-to-mass ratio,  which plays an important role in the classification of the solutions. 

As already noticed, the geometry described above is an exact solution of certain extensions of GR such as the Born-Infeld model \cite{ORGS14}  and quadratic gravity  \cite{Olmo:2012nx} coupled to an electric field. The reason why such solutions were not known previously is because we considered those theories from a metric-affine perspective, i.e., assuming that the metric and the affine connection are independent geometric entities, which yields field equations different from those found in the typical metric approach \cite{Olmo}. This formulation of modified gravity yields second-order field equations that recover Minkowski (or de Sitter) space-time in vacuum, yields no extra propagating degrees of freedom, and is ghost free. The matter source threading the corresponding space-time is an electrostatic spherically symmetric Maxwell field, which means that the classical energy conditions are all satisfied. We also note that this static geometry can be obtained by considering the dynamical process of collapse of charged radiation fluids in a Vaidya-type space-time \cite{lmor}.

From the above definitions, one can verify that the GR geometry is quickly recovered as soon as $|x|$ is slightly greater than $r_c$ (recovering $B(r)\approx A(r) \approx 1-r_S/r+r_q^2/(2r^2)$, and $r^2\approx x^2$) and also in the limit $l_\epsilon\to 0$. Thus, for scales larger than the wormhole size $r_c$,  this space-time looks much the same as the standard Reissner-Nordstr\"om solution of the Einstein-Maxwell field equations. As $x\to 0$, however, the area function $r^2(x)\approx r_c^2+x^2/2$ reaches a minimum, and $A(r)$ dramatically departs from GR, behaving as (from now on we set $l_\epsilon=l_{Planck}$ for notational convenience)

\begin{eqnarray}
A(r)&\approx & \left\{ \begin{array}{lr}
1-\frac{r_S}{r}+\frac{r_q^2}{2r^2} & \text{ if } |x|\gg r_c \\
\frac{N_q}{4N_c}\frac{\left(\delta _1-\delta _c\right) }{\delta _1 \delta _c }\sqrt{\frac{r_c}{ r-r_c} } +\frac{N_c-N_q}{2N_c}& \text{ if } x\to 0
                                 \end{array}\right. \ ,
\end{eqnarray}
where $N_q=|q/e|$ is the number of charges and $N_c\approx 16.55$. This behaviour turns the Kretschmann scalar from ${R^\alpha}_{\beta\gamma\lambda}{R_\alpha}^{\beta\gamma\lambda}\approx 14r_q^4/r^8$ in the Reissner-Nordstr\"om solution into ${R^\alpha}_{\beta\gamma\lambda}{R_\alpha}^{\beta\gamma\lambda}\approx  (\delta_1-\delta_c)^2K_2/(r-r_c)^3+(\delta_1-\delta_c)K_1/(r-r_c)^{3/2}+ K_0$ in our case, with the $K_i$ constants. 

If the charge-to-mass ratio $\delta_1$ takes the particular value $\delta_1=\delta_c$, the metric and all curvature scalars are regular everywhere and, naturally, all time-like and null geodesics smoothly extend across the wormhole. In this case, if $N_q>N_c$, which implies $\lim_{x\to 0} A(r)<0$, the wormhole is hidden in between two event horizons symmetrically located around $x=0$ ($r=r_c$), connecting in this way two Schwarzschild-like geometries. If $N_q<N_c$, there are no event horizons, and the wormhole connects two Minkowski-like spaces. When $\delta_1 \neq \delta_c$, however, curvature divergences arise at the throat and the geodesic equation must be studied in detail to determine if those configurations are geodesically complete or not.

\begin{figure}[h]
\includegraphics[width=0.45\textwidth]{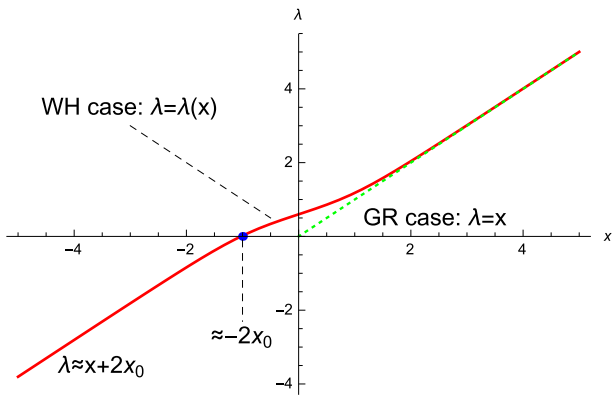}
\caption{Affine parameter $\lambda(x)$ as a function of the radial coordinate $x$ for radial null geodesics (outgoing in $x>0$). In the GR case (green dashed curve in the upper right quadrant), $\lambda=x$ is only defined for $x\ge 0$. In this plot $E=1$ and the horizontal axis is measured in units of $r_c$. } \label{Fig:affine_nullradial}
\end{figure}

In terms of an affine parameter $\lambda$, the geodesics of (\ref{eq:ds2}) satisfy \cite{ORGSPa} ($\kappa=0,1$ for null and time-like, respectively)
\begin{equation}\label{eq:dx/dl}
\frac{1}{\left(1+\frac{r_c^4}{r^4(x)}\right)^2}\left(\frac{dx}{d\lambda}\right)^2=E^2-A(r)\left(\kappa+\frac{L^2}{r^2(x)}\right) \ ,
\end{equation}
where $E$ and $L$ are constants which, in the time-like case, represent the total energy and angular momentum per unit mass. For null geodesics with $L=0$, we get
\begin{equation}\label{eq:nullradial0}
\frac{1}{\left(1+\frac{r_c^4}{r^4}\right)^2}\left(\frac{dx}{d\lambda}\right)^2=E^2 \ ,
\end{equation}
whose solution is plotted and compared with the GR prediction in Fig.\ref{Fig:affine_nullradial}. Given that (\ref{eq:nullradial0}) is independent of $\delta_1$, all configurations (with or without curvature divergences) have an identical behavior, which confirms the extendibility of null geodesics across $x=0$  in these space-times.

For null and time-like geodesics with $L\neq 0$, near the wormhole the effective potential
\begin{equation}\label{eq:Vgeo}
V(x)=A\left(\kappa+\frac{L^2}{r^2(x)}\right) \
\end{equation}
can be approximated as $V(x)\approx -a/|x|$, with $a=\left(\kappa+\frac{L^2}{r_c^2}\right)\frac{ N_q (\delta_c-\delta_1)r_c}{2N_c\delta_c \delta_1 }$. One readily sees that if $\delta_1>\delta_c$ (Reissner-Nordstr\"{o}m-like case), an infinite potential barrier arises which prevents any geodesic from reaching $x=0$. This behavior is also observed in GR, where such geodesics can never reach the singularity \cite{Chandra}. In the Schwarzschild-like case, $\delta_1<\delta_c$, the potential becomes infinitely attractive as $x\to 0$ and turns (\ref{eq:dx/dl}) into\begin{equation}
\frac{d\lambda}{dx}\approx \pm\frac{1}{2}\left|\frac{x}{a}\right|^{\frac{1}{2}}  \ \to \lambda(x)\approx   \pm\frac{x}{3}\left|\frac{x}{a}\right|^{\frac{1}{2}} \ ,
\end{equation}
where $\pm$ is for outgoing/ingoing geodesics. In the case of GR, time-like radial geodesics near the singularity behave as $\lambda(r)\approx\pm \frac{2}{3}r\left(\frac{r}{r_S}\right)^{\frac{1}{2}}$. Given that $r>0$, ingoing/outgoing geodesics in GR end/start at  $\lambda(0)=0$ and, therefore, are incomplete. In our case, on the contrary, the fact that $x\in]-\infty,+\infty[$ allows to smoothly extend the affine parameter $\lambda$ across $x=0$ to the whole real axis. It is also worth noting that the conclusion of the singularity theorems \cite{Theorems}, that in GR prevent the Schwarzschild geometry to be extended, does not hold for these wormhole space-times, because although the electromagnetic stress-energy tensor fulfills the weak energy condition, $T_{\mu \nu} k^\mu k^\nu \geq 0$ for $k^\mu$ non space-like, it does no longer imply $R_{\mu \nu}(g) k^\mu k^\nu \geq 0$ in our gravity theories.

\section{Congruences}

We have just seen that there are no restrictions for null, time-like or space-like geodesics to be extended across the wormhole (when they can reach it). Nonetheless, given that physical observers can be more accurately described as geodesic congruences, one should consider the impact of curvature divergences in the evolution of nearby geodesics. For this purpose, we consider the line element (\ref{eq:ds2}) written in coordinates adapted to an observer in free fall and study if causal contact between nearby time-like observers is lost at any time of the transit across the hole.  In freely falling coordinates, (\ref{eq:ds2}) turns into  \cite{Olmo:2016fuc}
\begin{equation}\label{eq:FF}
ds^2=-d\lambda^2+(u^y)^2d\xi^2  +r^2(\lambda,\xi)d\Omega^2 \ ,
\end{equation}
where $\lambda$ is the proper time (affine parameter), $\xi$ measures the {\it radial} separation between geodesics, and $u^y\equiv dy/d\lambda$, with $dy=dx/(1+r_c^4/r^4(x))$. Focusing on the Schwarzschild-like case, $\delta_1<\delta_c$, which is the only configuration in which time-like geodesics can go through the wormhole,  $(u^y)^2$ can be approximated as $(u^y)^2 \simeq a/|x| \simeq (\frac{3}{a} |\lambda-E\xi|)^{-\frac{2}{3}}$, which turns (\ref{eq:FF}) into
\begin{equation}\label{eq:FF_WH}
ds^2\approx-d\lambda^2+\left(\frac{3}{a} |\lambda-E\xi|\right)^{-2/3}d\xi^2\ .
\end{equation}
One should note that as the wormhole throat is approached, $\lambda-E\xi\to 0$,
the physical spatial distance between any two infinitesimally nearby geodesics diverges: $dl_{Phys}=\left(\frac{3}{a} |\lambda-E\xi|\right)^{-1/3}d\xi$ (see \cite{Olmo:2016fuc} for a more elaborate discussion of this point using Jacobi fields). This could be seen as an indication that strong tidal forces could rip apart any infalling body, thus signaling the presence of a {\it strong singularity}.  However, for any finite {\it comoving} separation $\Delta\xi\equiv \xi_1-\xi_0$, the physical spatial distance $l_{Phys}\equiv \int |u^y| d\xi$ is always finite,
\begin{equation}\label{eq:lphys}
l_{Phys}\approx \left(\frac{a}{3}\right)^{1/3}\frac{1}{E}\left||\lambda-E \xi_0|^{2/3}-|\lambda-E \xi_1|^{2/3}\right| \ ,
\end{equation}
which casts doubts on the fate of such bodies and the actual {\it strength} of those curvature divergences.

\begin{figure}[t]
\includegraphics[width=0.35\textwidth]{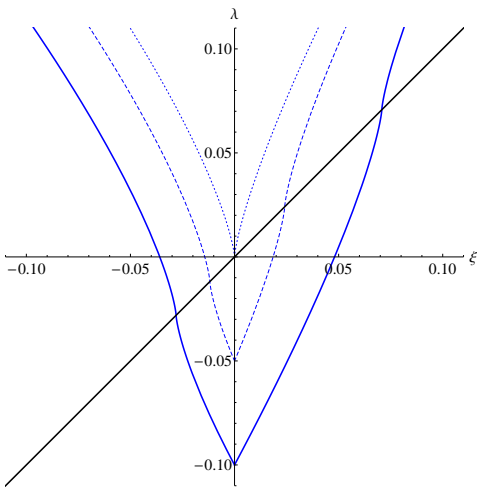}
\caption{Trajectories of light rays emitted by a freely falling observer from $\xi=0$ at different times shortly before reaching the wormhole throat in the Schwarzschild-like configuration. The rays going to the left/right represent ingoing/outgoing null geodesics. Given that the observer is inside an event horizon, both ingoing and outgoing light rays end up hitting the wormhole, which is located at the oblique (solid black) line $\lambda - E\xi= 0$ (in the plot $E=1, a=3$). }\label{fig:light-cones}
\end{figure}

It is thus necessary to clarify if the constituents making up an object that reaches the wormhole lose causal contact because of the divergent {\it spatial} stretching that affects its infinitesimal elements (infinite tidal forces). In such a case, the interactions that keep the object cohesioned would no longer be effective, resulting in disruption or disintegration of the body. However, as shown in Fig.\ref{fig:light-cones}, a fiducial observer at $\xi=0$ never loses causal contact with its neighbors. Indeed, in \cite{Olmo:2016fuc} we show that the proper time a light ray takes in a round trip from $\xi=0$ to any nearby geodesic is always finite as the wormhole is crossed. The divergent stretching of ({\it infinitesimal}) spatial distances is, therefore, not observable.  We can thus conclude that extended objects going through the wormhole, where curvature scalars generically diverge, do not undergo destructive deformations and, therefore, can effectively cross this apparently troublesome region.

\section{Scattering experiments}

As a third test to verify that curvature divergences do not affect the well-posedness of physical laws in our space-time (\ref{eq:ds2}), we consider the propagation of scalar waves in this background. According to the cosmic censorship conjecture, naked singularities should not exist in Nature. This idea is traditionally invoked in the literature to interpret naked singularities as unphysical artifacts of Einstein's theory. From this physical perspective, therefore, one can say that naked divergences are the worst-case scenario. From a technical viewpoint, this turns out to be the simplest situation because there exists a time-like Killing vector over the whole space, which facilitates the choice of coordinates and the separation of variables. Our approach is also valid for the region inside the inner horizon, where the wormhole is a time-like hypersurface. We will thus focus on configurations with $\delta_1>\delta_c$  (Reissner-Nordstr\"{o}m-like). Taking the scalar field equation $(\Box-m^2) \phi=0$, we decompose it in modes of the form $\phi_{\omega, l m}=e^{-i\omega t} Y_{lm}(\theta,\varphi) f_{\omega,l}(x)/r(x)$, where $Y_{lm}(\theta,\varphi)$ represents spherical harmonics. Using the radial coordinate $y=\int dx/A(1+r_c^4/r^4)$, the $f_{\omega,l}(x)$ are governed by a Schr\"{o}dinger-like equation of the form

\begin{equation}
-f_{yy}+V_{eff}f=\omega^2f,
\end{equation}
where
\begin{equation}\label{eq:Veff}
V_{eff}=\frac{r_{yy}}{r}+A(r)\left(m^2+\frac{l(l+1)}{r^2}\right) \ .
\end{equation}
This potential quickly converges to the GR prediction for $r\gg r_c$ but near $r=r_c$ diverges  as $V_{eff}\approx k/|y|^{1/2}$, where
\begin{equation}\label{eq:Veff}
k\equiv{\sqrt{\frac{(\delta_1-\delta_c)N_q }{\delta_1\delta_c N_c}}}\frac{(N_c[m^2r_c^2+1+l(l+1)]-N_q)}{N_c(8r_c^3)^{1/2}}
\end{equation}

Low frequency modes are almost entirely reflected by the centrifugal barrier in much the same way as it happens in GR (see \cite{Giveon:2004yg} for details). We thus focus on high-energy modes with sufficient energy to overcome this barrier and interact with the wormhole. The effective potential is divergent at the wormhole throat, but as is well known from the example of the Coulomb potential, this does not mean that the wave function is singular. Actually, the leading behaviour of the wave function $f_{\omega l} \sim y^\rho$ is given by the characteristic exponents $\rho=0,1$, which are perfectly regular. Considering the $m^2=0$ case for simplicity, it is easy to see that $V_{eff}$ may transit from an infinite well to an infinite barrier as $l$ increases if $N_q>N_c$. The divergent barrier is reminiscent of the repulsive potential  seen by geodesics in this same configurations. Using the mode decomposition, we thus define an incoming wave packet from past null infinity (in the naked case, or from the inner horizon in the black hole case) and study its interaction with the wormhole. As noticed above, the behavior will depend dramatically on the angular momentum of the incident mode.

With a notational change, the information contained in the parameters $l,N_q,\delta_1$ and the frequency $\omega$ can be condensed in a (dimensionless) parameter $\alpha$ defined as
\begin{equation}
y'=|k|^{\frac{2}{3}}y \ \rightarrow \ \alpha=|k|^{-\frac{2}{3}}\omega \ .
\end{equation}
The wave equation in the relevant region then becomes
\begin{equation}
f_{y'y'}+(\alpha^2\pm\frac{1}{\sqrt{y'}})f=0 \ ,
\end{equation}
with the $\pm$ sign set by the sign of $k$ (infinite well or barrier). Now we can compute numerically the transmission coefficient for a given value of $\alpha$. The full potential decays appropiately as the radius increases and the wave function can be approximated asymptotically as $A_\text{in}\exp(i\omega(t- r))+A_\text{out}\exp(i\omega(t+ r))$ on one side of the wormhole, and $B_\text{in}\exp(i\omega(t-r))+B_\text{out}\exp(i\omega(t+ r))$ on the other. To obtain the transmission coefficient, we give initial conditions $f$, $f^\prime$ at a point $x_\text{out}$ across the wormhole with the condition $B_\text{in}=0$, and evolve the wave function up to a point $x_\text{in}$ on the side of the wormhole where the wave was sent. From the values of $f(x_\text{in})$, $f^\prime(x_\text{in})$, $f(x_\text{out})$, $f^\prime(x_\text{out})$ we can extract the amplitudes $A_\text{in}$, $A_\text{out}$, $B_\text{out}$. The transmission coefficient is simply $T=|B_\text{out}/A_\text{in}|^2$. There is a a caveat if we use the approximated effective potential, because it is long-range and the wave function cannot be approximated with exponentials as before. We can use the WKB approximation far from the wormhole to obtain the amplitudes $A_\text{in}$, $A_\text{out}$, $B_\text{out}$, but this would introduce a phase respect to the amplitudes calculated using the full potential. However, the transmission coefficient would not be affected, and this is the method we have used.

\begin{figure}[t]
\includegraphics[width=0.35\textwidth]{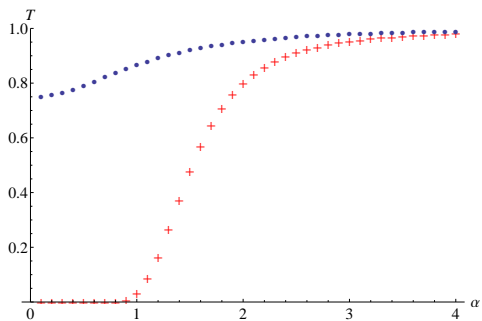}
\caption{Transmission coefficient for the potential well $-\frac{1}{\sqrt{|y'|}}$ (blue points) and the  barrier $+\frac{1}{\sqrt{y'}}$ (red crosses).}\label{TransAlpha}
\end{figure}

In Fig. \ref{TransAlpha} the transmission coefficient as a function of $\alpha$ is plotted (for the approximated $V_{eff}$).  For $k<0$ (infinite well), most of the wave is transmitted, with the transmission coefficient tending to $1$ as $\alpha$ grows. For $k>0$ (infinite barrier), a sigmoid profile arises like in typical barrier experiments. For $\alpha$ above $\sim 1.5$, the wave is transmitted almost entirely, and below that threshold it is almost completely reflected.

Consider now a case with constant $\omega$. Depending on the number of charges, $k$ can be either positive or negative for $l=0$. With growing $l$, $k$ grows as $\sim l^2$. The transmission for $k$ negative is very high, and approaches $1$ as $k\to 0$. When $k$ changes sign as $l$ grows, $\alpha\to \infty$ and the transmission is near $1$. For bigger $l$'s, $\alpha$ begins to decrease, and around the threshold $\alpha \sim 1.5$, characterized by $l=l_\text{max}$, the partial waves change from being almost totally transmitted to being almost totally reflected. A rough estimate of the cross section can thus be obtained by considering that the transmission is $1$ for partial waves with $l\leq l_\text{max}$, and $0$ for $l>l_\text{max}$. The transmission cross section would thus be

\begin{figure}[t]
\includegraphics[width=0.35\textwidth]{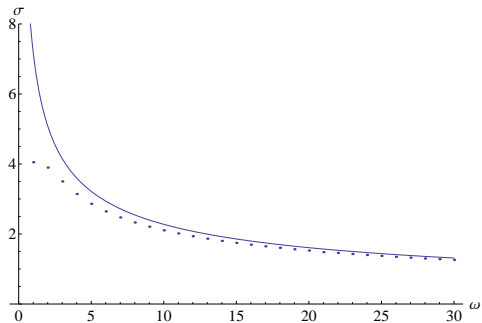}
\caption{Transmission cross section versus frequency for a naked divergence with $\delta_1=1$, $N_q=10$, calculated numerically (dots) for the exact $V_{eff}$. The continuous line shows the approximation $\sigma \propto \omega^{-\frac{1}{2}}$.}\label{CrossSection}
\end{figure}

\begin{equation}
 \sigma=\frac{\pi}{\omega^2}\sum_{l=0}^{l_\text{max}}(2l+1)1=\frac{\pi}{\omega^2}(1+l_\text{max})^2
\end{equation}
The variation of $\sigma$ in terms of $\omega$ is illustrated in Fig. \ref{CrossSection}. As $\omega \rightarrow \infty$, we have $l_\text{max} \propto \omega^{\frac{3}{4}}$, which confirms that $\sigma$ decreases  as $\sim \omega^{-\frac{1}{2}}$.

\section{Summary and conclusions}

Two fundamental properties typically demanded from quantum theories of gravity (often seen as one and the same)  are the removal of space-time singularities and the regularization of classical curvature divergences. We have shown here that only the former is physically relevant and can be achieved in classical geometric scenarios with nontrivial topology and satisfying all the energy conditions. Though our analysis has focused on a four-dimensional setting, our results are still valid in arbitrary dimensions, where exact analytical solutions with wormhole structure, and which are geodesically complete, have been found \cite{fan5} (we note that geodesic completeness via wormholes can also be achieved in certain metric-affine $f(R)$ theories \cite{fR}). Higher dimensional models confirm that the resolution of black hole singularities is a highly non-perturbative phenomenon, being the GR solution an excellent approximation down to the very throat of the wormhole, where the geometry quickly changes to account for this topological structure. We also note that the theories that generate the solution (\ref{eq:ds2}) also replace the big bang singularity by a cosmic bounce in a very robust manner \cite{Odintsov:2014yaa,Barragan:2010qb,Banados:2010ix}.

The generic absence of space-time singularities in these theories without requiring exotic matter-energy sources puts forward the existence of an important gap in our understanding of gravitation in metric-affine geometries. Interestingly, this type of geometries seem to play an important role in the description of continuum systems with a microstructure \cite{Lobo:2014nwa}, such as Bravais crystals or graphene. This provides further  motivation for their study in gravitational scenarios.

\section*{Acknowledgments}

G.J.O. is supported by a Ramon y Cajal contract, the Spanish grants FIS2014-57387-C3-1-P and FIS2011-29813-C02-02 from MINECO, the grants i-LINK0780 and i-COOPB20105 of the Spanish Research Council (CSIC), the Consolider Program CPANPHY-1205388, and the CNPq project No. 301137/2014-5 (Brazilian agency). D. R.-G. is funded by the Funda\c{c}\~ao para a Ci\^encia e a Tecnologia (FCT) postdoctoral fellowship No.~SFRH/BPD/102958/2014, the FCT research grant UID/FIS/04434/2013, and the NSFC grant No.~11450110403.

\end{document}